\documentclass[aps, showpacs,pra,twocolumn] {revtex4}
\usepackage{hyperref}
\usepackage{amsmath }
\usepackage{amssymb}
\usepackage{epsfig}
\usepackage{natbib}
\usepackage{bm}

\begin{document}
\bibliographystyle{revtex}

\title{ Cosmological Production of Fermions in a
Flat Friedman Universe
 \\
  with Linearly Growing Scale Factor: Exactly
Solvable Model }

\author{S.V. Anischenko}
\email{Lesavik@yandex.ru}
 \affiliation{Belarus State University, Bobruiskaya 5, Minsk 220050,
Belarus}
\author{S.L. Cherkas}
\email{cherkas@inp.minsk.by}
 \affiliation{Institute for Nuclear Problems, Bobruiskaya
11, Minsk 220050, Belarus}
\author{V.L. Kalashnikov}
\email{v.kalashnikov@tuwien.ac.at}
 \affiliation{Institut f\"{u}r Photonik, Technische
Universit\"{a}t Wien, Gusshausstrasse 27/387, Vienna A-1040,
Austria}

\received{ \today }

\begin{abstract}We consider an exactly solvable model for production of  fermions
in the Friedman flat universe with a scale factor linearly growing
with time. Exact solution expressed through the special functions
admit an analytical calculation of the number density of created
particles. We also discuss in  general the role of the phenomenon
of the cosmological particle production in the history of
universe.
\end{abstract}
\maketitle

\section{Introduction}
Although a cosmological particle production
\cite{Par69,SexUrb69,ZelSta71,frol} is well known phenomenon
\cite{birrel, grib,gavr}, exactly solvable models
\cite{BerDun,Kandr,Huang} are of interest because they would allow
comparing the different approximate calculation methods for the
number of produced particles. One of the problems in this field is
to define the in- and out- vacuum states if the metric has not an
asymptotic form corresponding to that of the Minkowski space-time
\cite{birrel,Fulling79}.

As the approximate methods, the adiabatic vacuum \cite{birrel},
WKB series \cite{Winitzky} and instantaneous Hamiltonian
diagonalization \cite{grib} methods are widely used. Evidently,
the later method does not define an exact vacuum state. As is
shown in Ref.\cite{vestn}, a certain superposition of the
 states, which diagonalize the instantaneous Hamiltonian, is needed
 to obtain the exact vacuum state.

The method has been suggested in Ref. \cite{nonlin}, which allows
finding the exact vacuum state by minimization of some functional.
Here, on basis of this method, we have found the exact vacuum
states for the case of a flat universe with a linearly growing
scale factor. Such a model has appeared in Ref. \cite{conf}, where
the universe driven by the vacuum has been considered. Although,
that model in its present version contradicts the nucleogenesis
theory, which insists on an existence of the radiation epoch, it
is of theoretical interest as a toy model. It should be noted,
that the scalar particle creation in such a Universe was
considered earlier \cite{hartle} (See e.g. Ref. \cite{jap}) on
basis of the WKB approximation, and father exact solution have
been obtained \cite{Huang}. For fermions energy-momentum tensor
have been calculated in Ref. \cite{var}.

 Let us also emphasize, that the
flat universe with a linearly growing scale factor differs
substantially from the Milne-like empty closed universe in a sense
that the later can be transformed to the region of the Minkowski
space-time \cite{flor} and does not exhibit a particle production.


\section{Fermion production}

Let us consider the fermion  in the expanding  universe. After
decomposition of the bispinor $\psi(\bm r,\eta)$ in the complete
set of modes
 $\psi(\bm
r)=\sum_{\bm k} \psi_{\bm k}(\eta)\, \mbox{e}^{\mbox{i} {\bm k}\bm
r}$, the Lagrangian of the fermion field in the expanding universe
\cite{2} takes the form

\begin{eqnarray}
 L=\sum_{\bm k}\frac{\mbox{i}\, a^3 }{2}\psi^+_{\bm
k}\partial_\eta\psi_{\bm k}-\frac{\mbox{i}\, a^3}{2}
\partial_\eta \psi^+_{\bm k}\psi_{\bm k}~~~~~~~~~~~~\nonumber\\-a^3\psi^+_{\bm k}
(\bm \alpha \bm k)\,\psi_{\bm k}-a^4\, m \psi^+_{\bm k}\beta
\psi_{\bm k},
\end{eqnarray}
where $\eta$ is the conformal time, $a(\eta)$ is the universe
scale factor.

The equation of motion is
\begin{equation}
i{\psi}^\prime_k- (\bm \alpha  \bm k){\psi}_{\bm k} +\mbox{i}
\frac{3a^\prime}{2a}{\psi}_{\bm k}-m\,a \beta{\psi}_{\bm k}=0,
\label{rr}
\end{equation}
The fermion field is quantized as
\begin{equation}
\hat \psi_{\bm k}={\hat b}^+_{-\bm k,s}{ v}_{-\bm k,s}+{\hat
a}_{\bm k,s}u_{\bm k,s},
\end{equation}
where the bispinor is:
\[u_{\bm k,s}(\eta) = \frac{\mbox{i}\chi_k^\prime+m a\chi_k}{a^{3/2}}\left (
\begin {array}{c}
 \varphi_s \\
\frac{\chi_k(\bm \sigma \bm k)}{i\chi_k^\prime+m a\chi_k}\varphi_s
\end {array}
\right)~,
\]
where $\varphi_s$ are
 $\varphi_+=\left (
\begin {array}{c}
1 \\
0
\end {array}
\right)$ and $\varphi_-=\left (
\begin {array}{c}
0 \\
1
\end {array}
\right)$.

The bispinor $v_{\bm k,s}$  is expressed as  $v_{\bm
k,s}=\mbox{i}\gamma^0\gamma^2(\bar u_{\bm k,s})^{T}$, where the
symbol $T$ denotes a transposed vector, $\bar u=u^+\gamma^0$ and
the representation of the Dirac matrices is the same as in Refs.
\cite{lan4,cher}. The functions $\chi_{k}(\eta)$ satisfy \cite{2}

\begin{eqnarray*}
\chi^{\prime\prime}_{k }+ \left(k^2+m^2 a^2-\mbox{i} m
a'\right)\chi_{ k}=0, \label{1}
\\
 k^2\chi_{ k}  \chi^*_{ k} +\left(a m \chi^*_{k}-\mbox{i} {\chi
^\prime_{k}}^*\right) \left(a m \chi_{k} +\mbox{i} \chi^\prime_{
k} \right)=1, \label{rel1}
\end{eqnarray*}
that corresponds to a time-dependent oscillator with a complex
frequency.

Linearly growing scale factor

\begin{equation}
a(t)=a_0 H_0 t \label{scalef}
\end{equation}
 corresponds to the exponential
dependence $a(\eta)=a_0\exp(\mathcal H \eta)$ in the conformal
time $d\eta=dt/a$, where the conformal Hubble constant is
$\mathcal H=a_0 H_0$. To describe the particles production, one
should define the in- and out- vacuum states.
 There is a lot of
solutions connected by the Bogoliubov transformation \cite{bog}.

For the fermions with the momentum $\bm k$ directed along
$z$-axes, the Bogoliubov transformations  are:
\begin{eqnarray}
{\mathcal U}_+(k)=\cos r_k \,{ u}_+(k)-\sin r_k e^{i\delta_k}\,{
v}_-(-k),\nonumber\\
{\mathcal V}_+(k)=\cos r_k \,{ v}_+(k)+\sin r_k e^{-i\delta_k}\,{
u}_-(-k),\nonumber\\
{\mathcal U}_-(k)=\cos r_k \,{u}_-(k)-\sin r_k e^{i\delta_k}\,{
v}_+(-k),\nonumber\\
{\mathcal V}_-(k)=\cos r_k \,{v}_-(k)+\sin r_k
e^{-i\delta_k}\,{u}_+(-k), \label{bg1}
\end{eqnarray}
where $u_\pm(k)$, $v_\pm(k)$, $\mathcal U_\pm(k)$, $\mathcal
V_\pm(k)$ denote $u_{\pm,\bm k}$, $v_{\pm,\bm k}$, $\mathcal
U_{\pm,\bm k}$, $\mathcal V_{\pm,\bm k}$ with $\bm k$ directed
along $z$-axes.   From Eqns. \eqref{bg1} it follows that the
functions $\chi_k$, $\mathcal X_k$ corresponding to the different
vacuums are connected as
\begin{eqnarray*}
{\mathcal X}_k=\cos r_k\, \chi_k(\eta)-e^{i\delta_k}\sin r
_k\left({m
a(\eta)}\chi^*_k(\eta)-i\chi_k^{*\prime}(\eta)\right)/k,\\
{\mathcal X}_k^*=\cos r_k\, \chi_k^*(\eta)-e^{-i\delta_k}\sin r
_k\left({m a(\eta)}\chi_k(\eta)+i\chi_k^{\prime}(\eta)\right)/k.
\end{eqnarray*}

 According to \cite{nonlin}, the quantity \begin{equation}
\sigma_k(\eta)=\frac{1}{2}\bigl( \chi^\prime_k(\eta)
\chi^{*}_k(\eta)+{ \chi}^{*\prime}_k(\eta)\chi_k(\eta)\bigr)
\label{sig10}
\end{equation}
is monotonic in the past but has oscillating behavior in the
future for the in-vacuum. For the out-vacuum state, it is
oscillating in the past but is monotonic in the future.

\begin{figure*}[t]
\hspace {-0.5 cm} \epsfxsize =13. cm \epsfbox{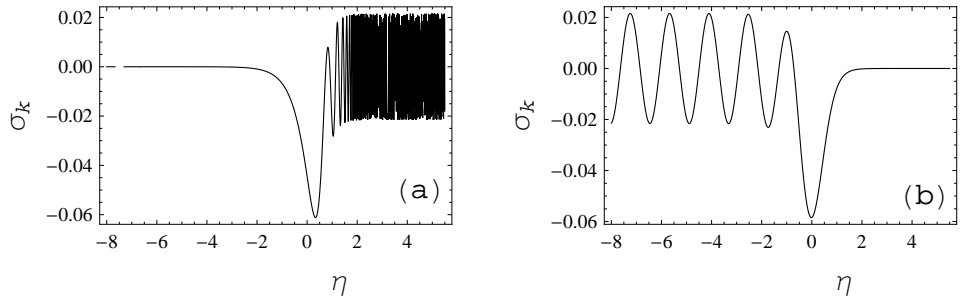}
\caption{(a) Dependence of the function $\sigma_k$ on the
conformal time for the in-vacuum state and (b) for the out-vacuum
state; $m=1$, $k=2$, $\mathcal H=2$.}
 \vspace{-0.5cm}
\label{s}
\end{figure*}

With the help of the functional minimization method \cite{nonlin},
we obtain that
\begin{eqnarray}
\chi_k=\sqrt{\pi } \sqrt{\frac{\mbox{csch}\left({2 \pi
k}/{\mathcal H}\right)}{\mathcal H k}} \exp\left({-{i m\,a_0
e^{\mathcal H \eta}}/{\mathcal H}-i k \eta}\right)~\nonumber\\
L_{{i k/{\mathcal H }}-1}^{(-{2 i k}/{\mathcal H})}\left({2 i
m\,a_0
   e^{\mathcal H \eta}}/{\mathcal H}\right)~
\end{eqnarray}
\noindent for the in-vacuum state and function $\sigma$ shown in
Fig. \ref{s}. (a) has the form

\begin{widetext}
\begin{equation}
\sigma_k=-\frac{\pi  m a_0 }{4 \mathcal H k}e^{\mathcal H \eta }
\text{sech}\left(\frac{\pi k}{\mathcal H}\right)
\biggl(J_{\frac{1}{2}-\frac{i k}{\mathcal
H}}\left(\frac{e^{\mathcal H \eta }
   m a_0}{\mathcal H}\right) J_{\frac{i k}{H}-\frac{1}{2}}\left(\frac{e^{\mathcal H \eta } m a_0}
   {\mathcal H}\right)+J_{-\frac{i k}{\mathcal H}-\frac{1}{2}}\left(\frac{e^{\mathcal H
   \eta } m a_0}{\mathcal H}\right) J_{\frac{i k}{\mathcal H}+\frac{1}{2}}\left(\frac{e^{\mathcal H \eta } m a_0}{\mathcal
   H}\right)\biggr).\label{expsig}
\end{equation}
\end{widetext}

For the out-vacuum state (see Fig. \ref{s}. (b))
\begin{eqnarray}
\mathcal X_k=\frac{1}{\mathcal H}\exp\left({{\pi  k}/{2\mathcal
H}-{i m\,a_0 e^{\mathcal H \eta}}/{\mathcal H}-i k \eta}
\right)~~~\nonumber\\U\left(-\frac{i k-\mathcal H}{\mathcal
H},1-\frac{2 i k}{\mathcal H},\frac{2 i e^{\mathcal H \eta}
m\,a_0}{\mathcal H}\right).
\end{eqnarray}
\noindent Here $U(a,b,c)$ is the confluent hypergeometric Kummer's
function and $L_m^{(\nu)}(z)$ is the generalized Laguerre function
\cite{ab}.

 Square of $\sin
r_k$ can be expressed as
\begin{widetext}
\begin{eqnarray}
\sin^2 r_k=\left|\frac{k \left(\chi_k  {\mathcal X}_k'-{\mathcal
X}_k \chi_k '\right)}{i m^2 a^2 \chi_k \chi_k^*-m a \chi_k^*
   \chi_k '+m a\chi_k \chi_k^{*'}+i k^2 \chi_k  \chi_k^*+i \chi_k
   '
   \chi_k^{*'}}\right|^{\,2}=\frac{1}{1+\exp\left({2\pi k/{\mathcal
   H}}\right)}.
\label{res1}
\end{eqnarray}
\end{widetext}

The density of the produced fermion is
\begin{eqnarray}
n=\frac{N}{V a^3}= \frac{2}{V
a^3}\int_0^\infty\sin^2r_k\,\frac{Vd^3{\bm
k}}{(2\pi)^3}\approx\frac{3}{40 \pi^5}\frac{\mathcal
H^3}{a^3}~~\nonumber\\=\frac{3}{40 \pi^5}H^3(t). \label{dens}
\end{eqnarray}

\section{Discussion and Conclusion}

It is interesting to compare the number of the produced fermions
with the result of Refs. \cite{grib,mam}, where the dependence
\begin{eqnarray}
a(t)=a_0\,t^q=\tilde a_0\eta^p, \label{aq}
\\~~~~\tilde a_0=a_0^{1/(1-q)}(1-q)^{q/(1-q)},\nonumber\\
p=q/(1-q),~~~~~~ 0<q<1
\end{eqnarray}
was considered. The density of the produced fermions as well as
scalar particles have been estimated as \cite{grib,mam}
\begin{equation}
n\sim m^3(m t)^{-3q}. \label{nn}
\end{equation} Under $q=1$ one has $n\sim
t^{-3}$. Up to a numerical multiplier this coincides with our
exact result given by (\ref{scalef}), (\ref{res1}), (\ref{dens}).
The numerical multiplier equals approximately to $\frac{3}{40
\pi^5}\approx 2.5\times 10^{-4}$.

Now we want to discuss the moment of time when the particles of
some mass $m$ are created. According to Ref. \cite{nonlin}, the
moment of time, from which the function $\sigma$ given by
\eqref{sig10} acquires oscillating behavior corresponds to the
particle creation.

From the analysis of the expression \eqref{expsig} it follows that
the particles of mass $m$ are created by time $\eta_1$ determined
by the equation $m a_0\exp\left(\mathcal H \eta_1\right)/\mathcal
H\approx 1$ that is $m/H(t_1)\approx 1$. By that moment of time
according to \eqref{dens} total density of a particles
approximately equals $n_1\sim H(t_1)^3$ (or equally $n_1\sim m^3$
) and further it is simply reduces with time as the $n(t)\sim n_1
\frac{a_1^3}{a^3(t)}$. From dimension arguments it can be argued
that for arbitrary $a(t)$ density number of the created particles
can be estimated as
\begin{equation}
n(t)\sim m^3\,\frac{a^3(t_1)}{a^3(t)}, ~~~~~~~~~H(t_1)\approx m.
\label{form}
\end{equation}
Let us check this formula for the dependence given by \eqref{aq}.
For the time of particle creation we have $ t_1\approx\frac{q}{m},
$ and finally come to \eqref{nn}.

It is also interesting to discuss in general the possible role of
the cosmological particle production in the creation of matter in
the Universe. According to the modern view decelerating fast-roll
stage preceded the inflationary stage \cite{vega}. At this stage
Hubble constant decreased down to the value $H(t_1)$ right up to
the time $t_1$ when the inflation begins. Let us estimate amount
of matter created before the inflation stage. At the inflation
stage typical value of the Hubble constant is $H_1=m$, where $m$
is the inflanton mass. It is an order of $m\sim 10^{-6}M_p$, where
$M_p\sim 10^{18}$ GeV is the Plank mass.
  Density of the inflantons  produced under the
 background of the coherent inflanton field is give by
\eqref{form}. During inflation and further expansion their density
is reduced in $a_1^3/a_0^3$ times where $a_0$ is the scale factor
of the present universe. Present time matter density created due
to this effect can be estimated as
\begin{equation}
\rho=m \,n_1\frac{a_1^3}{a_0^3}=(M_p \,10^{-6})^4\,
\frac{a_1^3}{a_0^3},
\end{equation}
where for simplicity we suggest that the created inflantons decay
to the massive particles, so that their general amount of mass is
conserved. Comparing this quantity with the observed density of
matter in the universe $\rho_c=M_p^2\, H_0^2=M_p^4 10^{-122}$ we
find that the expansion rate has to be $\frac{a_0}{a_1}=10^{33}$
in order to the densities be equal.

That is the matter arising due to cosmological creation of
inflantons will be negligible only if the expansion rate of the
universe during inflation and further expansion  exceeds $10^{33}$
(76 e-foldings).

To summarize, we have presented the new exactly solvable model for
fermion production in the flat universe with a scale factor
linearly growing with time. Spectrum of the produced  fermions
over physical momentums $p=k/a$
\[n(p)=\frac{1}{ \exp(2\pi p/ H(t))+1}
\]
is thermal at large $p$, and the Hubble constant divided by $2\pi$
plays a role of the temperature. The number density of the created
particles does not depend on the particle mass, and particles of
large mass can be easy created. However, the function $\sigma_k$
begins to oscillate later in the time, when the mass decreases.

The oscillations of this function mean that the real particles
appear \cite{nonlin}. Thus, in the limit of $m\rightarrow 0$, the
particles would be created infinitely late.

\addcontentsline{toc}{section}{References}

\end{document}